\documentstyle[prd,aps]{revtex}

\begin{document}

\newtheorem{thm}{Theorem}[subsection]
\newtheorem{prop}[thm]{Proposition}
\renewcommand{\topfraction}{1.0}
\twocolumn[\hsize\textwidth\columnwidth\hsize\csname
@twocolumnfalse\endcsname

\title{Solution generating  in scalar-tensor
theories  with a massless scalar field and  stiff perfect fluid
as a source}
%{\small}}

\author{Stoytcho S. Yazadjiev }

\address{Department of Theoretical Physics, Faculty of Physics, Sofia
University,\\
5 James Bourchier Boulevard, 1164 Sofia, Bulgaria \\ {\tt E-mail:
yazad@phys.uni-sofia.bg}}

 \maketitle
\bigskip

\maketitle

\begin{abstract}
We present a method for generating solutions in some scalar-tensor
theories with a minimally coupled massless scalar field or
irrotational stiff perfect fluid as a source. The method is based
on the group of  symmetries of the dilaton-matter sector in the
Einstein frame. In the case of Barker's theory the dilaton-matter
sector possesses $SU(2)$ group of symmetries. In the case of
Brans-Dicke and the theory with "conformal coupling", the dilaton-
matter sector has $SL(2,R)$ as a group of symmetries. We describe
an explicit algorithm for generating exact scalar-tensor solutions
from solutions of Einstein-minimally-coupled-scalar-field
equations by employing the nonlinear action of the symmetry group
of the dilaton-matter sector. In the general case, when the
Einstein frame dilaton-matter sector may not possess nontrivial
symmetries we also present a solution generating technique which
allows us to construct exact scalar-tensor solutions starting
with the solutions of Einstein-minimally-coupled-scalar-field
equations. As an illustration of the general techniques, examples
of explicit exact solutions are constructed. In particular, we
construct inhomogeneous cosmological scalar-tensor solutions whose
curvature invariants are everywhere regular in space-time. A
generalization of the method for scalar-tensor-Maxwell gravity is
outlined.
\end{abstract}

\vskip 0.8cm


\vskip2pc]

%\narrowtext

\section{Introduction}

Scalar-tensor theories of gravity are  considered as the most
natural generalizations of general relativity
\cite{BD}-\cite{KW2}. In these theories gravity is mediated not
only by the metric of space-time but also by a scalar field (the
so called gravitational scalar). Scalar-tensor theories contain
arbitrary functions of the scalar field that determine the
gravitational "constant" as a dynamical variable. From a
theoretical point of view it should be noted that specific
scalar-tensor theories arise naturally as  a low-energy limit of
string theory.

In the weak field limit scalar-tensor theories differ slightly
from general relativity. In the strong field regime, however, the
predictions of  scalar-tensor theories may differ drastically
from those of general relativity as it was shown in Refs.
\cite{DEF2} and \cite{DEF3}.

Scalar-tensor theories have also attracted much  interest in
cosmology (see Refs.\cite{LS}-\cite{M} and references therein).

The progress in the understanding of scalar-tensor theories of
gravity is closely connected with finding and investigation of
exact solutions. A theoretical discussions of many aspects of the
early universe, gravitational waves, gravitational collapse and
the structure of the compact objects within the framework of
scalar-tensor theories (as in general relativity) necessitates
the use of exact solutions. Besides the theoretical motivation
for construction of exact solutions, there is also a more
practical one. With the advent of numerical calculations, exact
solutions are useful as comparisons for numerical  and
approximate solutions and as checks of the computer codes.

Solving of  scalar-tensor theories equations in the presence of a
source is a difficult task due to their complexity in the general
case. In fact, scalar-tensor gravity equations are much more
complicated than  Einstein equations. In the so called Einstein
frame the sourceless scalar-tensor equations are reduced to
Einstein equations with a minimally coupled scalar field. In this
case much progress has been achieved in finding exact homogeneous
and inhomogeneous cosomological solutions \cite{C}-\cite{WIM}.
Little has been done in solving scalar-tensor equations with a
source. The known solutions are perfect fluid  homogeneous
cosmological solutions depending on the time coordinate only (see
Refs.\cite{B}-\cite{Coley} and references therein). In
Ref.\cite{B}, Barrow investigated a method which enables exact
solutions to be found for vacuum and radiation dominated
Friedmann-Robertson-Walker (FRW) universes of all curvatures in
scalar-tensor theories with an arbitrary form of the coupling
function $\omega(\Phi)$. Particular classes of solutions were
presented for specific choices of $\omega(\Phi)$, including
Brans-Dicke, Barker and Bekenstein theories. Barrow and Mimoso
\cite{BMIM} presented a method for deriving exact solutions for
flat FRW cosmological models with a perfect fluid satisfying the
equation of state $p=(\gamma -1)\rho$ where $\gamma$ is constant
with $0\le\gamma \le 2$, in scalar-tensor theories with an
arbitrary form for the coupling  function $\omega(\Phi)$. A
number of explicit solutions for inflationary universes and $p=0$
universes were obtained. Exact FRW cosmological solutions in
general scalar-tensor theories for stiff perfect fluid or
radiation were derived by Mimoso and Wands in Ref.\cite{MW1}.
Homogeneous but anisotropic cosmologies in scalar-tensor theories
of gravity were examined by Mimiso and Wands in Ref.\cite{MW2}.
The authors presented a method for deriving solutions for any
isotropic perfect fluid with a barotropic equation of state in a
spacially flat cosmology. For stiff fluid or radiation or in
vacuum the authors were able to obtain solutions in a number of
anisotropic Bianchi and Kantowski-Sachs metrics. Extending the
earlier work of \cite{BM}, \cite{B} and  \cite{BMIM}, Barrow and
Parsons \cite{BPAR} provided a detailed analysis of FRW universes
in a wide range of scalar-tensor theories of gravity. The authors
constructed a range of exact solutions for open, closed and flat
isotropic universes containing matter with an equation of state
$p\le (1/3)\rho$ and in vacuum. The early and late-time behaviours
of the solutions were examined, too. In Ref.\cite{BC}, Billyard
and Coley discussed the formal equivalences between Kaluza-Klein
gravity, Brans-Dicke theory and general relativity coupled to a
massless scalar field. Using the formal equivalences the authors
showed that exact solutions obtained in one theory will
correspond to analogous solutions in the other two theories. A
phase-space analysis of the FRW models in Brans-Dicke theory was
performed by Kolitch and Eardley  \cite{KE}. Their analysis was
improved on by Holden and Wands \cite{HW} who presented all FRW
models in a single phase plane. Particular attention was focused
on the early and late-time behaviour of the solutions and on
whether inflation occurs. The qualitative properties of
scalar-tensor theories of gravity were also studied by Coley in
Ref.\cite{Coley}. The author presented exact solutions which are
analogues of the general relativistic Jacobs stiff perfect fluid
solutions and vacuum plane wave solutions which act as past and
future attractors in the class of spatially homogeneous models in
Brans-Dicke theory.

It should be noted the the methods developed in \cite{B},
\cite{BMIM}, \cite{MW1}, and \cite{MW2} are solution generating
methods only for homogeneous cosmological solutions and are not
applicable to more general cases.

The purpose of this paper is to present a general  method for
generating exact solutions to the gravity  equations with a
minimally coupled massless scalar field (MCSF) and  irrotational
stiff perfect fluid within the framework of some scalar-tensor
theories whose Einstein frame dilaton-matter sector has nontrivial
symmetries. In the general case, when the Einstein frame
dilaton-matter sector may not possess nontrivial symmetries we
also present a general solution generating technique which allows
us to construct exact scalar-tensor solutions starting with the
solutions of Einstein-minimally-coupled-scalar-field (EMCSF). This
technique is based on the geodesics of the Riemannian metric
associated with the dilaton-matter sector.

The motivations to consider a MCSF as a source are the following.
In view of the complexity of the equations of the scalar-tensor
gravity, it is natural as a first step to consider a simple
source. On the other hand, in different contexts, the scalar
field (different from the gravitational scalar) plays an
important role in modern physics: the scalar field has been
proposed as a candidate for gravitational lensing
\cite{VNC},\cite{MB} and for dark matter at galaxies scales
\cite{GMV}, as well as at cosmological scales
\cite{CK},\cite{MUL},\cite{BP}.

Examples of different kinds of explicit exact solutions are also
considered.

\section{Scalar-tensor theories with a minimally coupled scalar
field as a source and  symmetries of dilaton - matter  sector}

Scalar-tensor theories are described by the following action in
the Jordan frame:
\begin{eqnarray} \label{JFA}
S = {1\over 16\pi G_{*}} \int d^4x \sqrt{-{\tilde
g}}\left({F(\Phi)\tilde R} - Z(\Phi){\tilde
g}^{\mu\nu}\partial_{\mu}\Phi
\partial_{\nu}\Phi  \right. \nonumber  \\ \left. -2 U(\Phi) \right) +
S_{m}\left[\Psi_{m};{\tilde g}_{\mu\nu}\right] .
\end{eqnarray}

Here, $G_{*}$ is the bare gravitational constant, ${\tilde R}$ is
the Ricci scalar curvature with respect to the space-time metric
${\tilde g}_{\mu\nu}$. The dynamics of the scalar field $\Phi$
depends on the functions $F(\Phi)$, $Z(\Phi)$ and $U(\Phi)$. In
order for the gravitons  to carry positive energy the function
$F(\Phi)$ must be positive. The action of matter depends on the
material fields $\Psi_{m}$ and the space-time metric ${\tilde
g}_{\mu\nu}$ but does not involve the scalar field $\Phi$ in order
for the weak equivalence principle to be satisfied. It should be
mentioned that the most used parametrization in the literature is
Brans-Dicke one, corresponding to $F(\Phi)=\Phi$ and $Z(\Phi)=
\omega(\Phi)/\Phi$.

It is much clearer to analyze the equations in the so-called
Einstein frame. Let us introduce the new variables $g_{\mu\nu}$
and $\varphi$, and define

\begin{eqnarray}\label{GSD}
g_{\mu\nu} = F(\Phi){\tilde g}_{\mu\nu} \,\,\, , \nonumber  \\
 \left(d\varphi \over d\Phi \right)^2 = {3\over
4}\left({d\ln(F(\Phi))\over d\Phi } \right)^2 + {Z(\Phi)\over 2
F(\Phi)} \,\,\, , \\
 {\cal A}(\varphi) = F^{-1/2}(\Phi) \,\,\, ,\nonumber \\
2V(\varphi) = U(\Phi)F^{-2}(\Phi) \,\,\, .\nonumber
\end{eqnarray}

From now on we will refer to $g_{\mu\nu}$ and $\varphi$ as
Einstein frame metric and dilaton field correspondingly.

In the Einstein frame the action (\ref{JFA}) takes the form

\begin{eqnarray}
S= {1\over 16\pi G_{*}}\int d^4x \sqrt{-g} \left(R -
2g^{\mu\nu}\partial_{\mu}\varphi \partial_{\nu}\varphi -
4V(\varphi)\right) \nonumber \\ + S_{m}[\Psi_{m}; {\cal
A}^{2}(\varphi)g_{\mu\nu}]
\end{eqnarray}

where $R$ is the Ricci scalar curvature with respect to the
Einstein frame metric $g_{\mu\nu}$.

We consider scalar-tensor theories with a minimally coupled
(massless) scalar filed (MCSF), $\sigma$, as a matter source. A
minimally coupled massless scalar field also corresponds to an
irrotational stiff perfect fluid with energy density

\begin{equation}
8\pi G_{*}{\tilde \rho}= 8\pi G_{*}{\tilde p} = - {\tilde
g}^{\mu\nu}\partial_{\mu}\sigma \partial_{\nu}\sigma
\end{equation}

and velocity field

\begin{equation}
{\tilde u}_{\mu} = \partial_{\mu}\sigma /\sqrt{-{\tilde
g}^{\mu\nu}\partial_{\mu}\sigma \partial_{\nu}\sigma}
\end{equation}

provided ${\tilde g}^{\mu\nu}\partial_{\mu}\sigma
\partial_{\nu}\sigma < 0$ \cite{TT}, \cite{WIM}.

The Jordan frame action for the scalar field is

\begin{equation}
S_{m}= {1\over 16\pi G_{*}}\int d^4x \sqrt{-{\tilde g}}\left(-2
{\tilde g}^{\mu\nu}\partial_{\mu}\sigma
\partial_{\nu}\sigma \right).
\end{equation}

In what follows we will consider only the special form of the
potential $U(\Phi)$: $U(\Phi)= 2\Lambda F^2(\Phi)$ or $V(\varphi)=
\Lambda=const$.

The full Einstein frame action then is

\begin{equation} \label{EFSFA}
S = {1\over 16\pi G_{*}} \int d^4x \sqrt{-g}\left(R - \Lambda  +
{\cal L}_{DM} \right)
\end{equation}

where

\begin{equation}\label{EFDMSL}
{\cal L}_{DM}= -2g^{\mu\nu}\partial_{\mu}\varphi
\partial_{\nu}\varphi  - 2{\cal A}^2(\varphi)
g^{\mu\nu}\partial_{\mu}\sigma \partial_{\nu}\sigma
\end{equation}

is the dilaton-matter sector of the theory Lagrangian.

Remarkably, in the case of some specific scalar-tensor theories
the dilaton matter-sector (\ref{EFDMSL}) of the theory possesses
hidden symmetries which allow us to generate exact solutions. In
order to unveil these symmetries we define a two-dimensional
abstract Riemannian space with a metric

\begin{equation}\label{DMSM}
dl^2 = d\varphi^2 + {\cal A}^2(\varphi)d\sigma^2 .
\end{equation}

Solution generation techniques consist in finding invariant
transformations of the dilaton-matter sector of the Lagrangian in
the action (\ref{EFSFA}). This is equivalent to finding the
isometry group of the metric (\ref{DMSM}). In two dimensions the
isometry group of (\ref{DMSM}) can be either $G_{1}$ or $G_{3}$.
Clearly our metric possesses $G_{1}$ isometry corresponding to
the Killing vector $\partial/\partial\sigma$. From a physical
point of view, however, this symmetry is not interesting because
it generates just a shift of the scalar field $\sigma$:
$\sigma\rightarrow \sigma + const$. The metric (\ref{DMSM}) has
$G_{3}$ group of isometries only when the Gauss curvature $K$ is
constant. Reversely, the constant curvature condition imposes a
differential equation for the function ${\cal A}(\varphi)$:

\begin{equation}\label{KDE}
K  = - {\cal A}^{-1}(\varphi){d^2 {\cal A}(\varphi)\over
d\varphi^2}.
\end{equation}

In this way, solving the diferential  equation (\ref{KDE}) we
obtain the scalar-tensor theories whose dilaton-matter sector
possesses group of isometries $G_{3}$.

When the Gauss curvature is positive ($K>0$) the group of
isometries is $SU(2)$ (see below). The scalar-tensor theories
corresponding to this case are characterized by $${\cal
A}(\varphi)= a\cos(\sqrt{K}\varphi + b)$$

which corresponds to the functions $F(\Phi)= \Phi$ and
$$Z(\Phi) = { (1 + 3K) - 3Ka^2\Phi\over 2K\Phi(a^2\Phi -1)}$$

where $a>0$ and $b$ are arbitrary constants. Since eq.(\ref{GSD})
define $\varphi$ up to a constant we may put $b=0$. In the case
$K=1$ and $a=1$ we obtain the Barker's theory \cite{Barker}.

For negative Gauss curvature ($K<0$) the isometry group is
$SL(2,R)$. The scalar-tensor theories whose dilaton-matter sector
has negative curvature are characterized by
$${\cal A}(\varphi) = a\exp(\sqrt{\mid K\mid} \varphi)+ b\exp(-\sqrt{\mid
K\mid}\varphi)$$ corresponding to the functions $F(\Phi)= \Phi$
and
$$ Z(\Phi)={(1-3\mid K\mid) + 12\mid K\mid ab \Phi \over 2\mid
K\mid \Phi(1-4ab\Phi)}$$

where $a>0$ and $b\ge 0$ are arbitrary constants. In the case
$b=0$ we obtain the Brans-Dicke theory with $\mid K \mid = 1/(3 +
2\omega )$\footnote{Here we consider the case $\omega > -3/2$.}.
When $ab\ne 0$ the function ${\cal A}(\varphi)$ can be presented
in the form

$${\cal A}(\varphi) = \Omega\cosh(\sqrt{\mid K\mid}\varphi +
c)$$

where $\Omega=2\sqrt{ab}$ and  $c$ is a constant. Without loss of
generality we may put $c=0$. In the special case $\Omega=1$ and
$\mid K\mid= 1/3$ we obtain the theory with "conformal coupling".
This theory is also described by the functions $F(\Phi)= 1-{1\over
6}\Phi^2$ and $Z(\Phi)=1$.

The scalar-tensor theories possessing flat ($K=0$) dilaton-matter
sector have an isometry group $Iso(\mathbf{R}^2)$ and are
described by \footnote{We have put the second constant in ${\cal
A}(\varphi)$ equal to one.}
$${\cal A}(\varphi) = 1 + a\varphi$$ corresponding to the
functions $F(\Phi)= \Phi$ and
$$Z(\Phi)= {1 -3a^2\Phi \over 2a^2\Phi^2}$$
where $a\ne 0$ is an arbitrary constant.

It is worth noting that he scalar-tensor theories with the same
group of isometries of the dilaton-matter sector  may have rather
different behaviour from a physical point of view. As for example
we may consider the Barker's theory with ${\cal A}(\varphi)\sim
cos(\varphi)$ and a theory with ${\cal A}(\varphi)\sim
cos(\sqrt{K}\varphi)$ where $K\ne 1$. For the Barker's theory the
effective gravitational "constant" is a real constant $G_{eff}\sim
G_{*}$ (to first order of the weak filed limit) while for a
theory with $K\ne 1$ the effective gravitational "constant" may
vary : $G_{eff}\sim cos^2(\sqrt{K}\varphi) + K
\sin^2(\sqrt{K}\varphi)$.

Below we consider in details  the symmetries of the
dilaton-matter sector for the scalar-tensor theories described
above and present solutions generating formulas. The general case
when dilaton-matter sector does not possess nontrivial symmetries
is also considered.

\subsection{Barker's theory}

Barker's theory \cite{Barker} is described by the functions
$F(\Phi)=\Phi$ and $Z(\Phi)= (4 - 3\Phi)/ \Phi (2\Phi - 2)$
corresponding to ${\cal A}^2(\varphi)= \cos^2(\varphi)$. The
metric (\ref{DMSM}) then is

\begin{equation}
dl^2 = d\varphi^2 + \cos^2(\varphi) d\sigma^2 \,\,\, .
\end{equation}

This metric can be considered as the standard metric on the unit
2-sphere. It is more convenient to present the metric in the well
known complex form. To do so we introduce the complex field

\begin{equation}
z= \cot\left(\varphi/2 + \pi/4 \right)e^{i\sigma} \,\,\, .
\end{equation}

We obtain then

\begin{equation}
dl^2 = 4 {dzd{\bar z}\over (1 + z{\bar z})^2 } \,\,\, .
\end{equation}

The metric is invariant under the transformations

\begin{equation}\label{SUT}
z \rightarrow z^{{\,}\prime} = {az + b \over - {\bar b}z + {\bar
a}}
\end{equation}

where
$$U = \pmatrix{a & b \cr -{\bar b} & {\bar a} } \in SU(2).$$

There is also an independent discrete symmetry $z\rightarrow
{\bar z}$ which corresponds to $\sigma \rightarrow - \sigma$.

We note that the $SU(2)$ transformations act nonlinearly on the
scalar fields but leave the Einstein frame metric invariant:
$g_{\mu\nu}^{{\,}\prime}= g_{\mu\nu}$. The Jordan frame metric
${\tilde g}_{\mu\nu}$, therefore, transforms under (\ref{SUT}) as

$${\tilde g}_{\mu\nu}^{{\,}\prime} \Phi^{{\,}\prime}=
{\tilde g}_{\mu\nu}\Phi .$$

The symmetries of the dilaton-matter sector can be used to
generate new solutions from known ones. In the case of Barker's
theory, any solution of Einstein equations with a minimally
coupled scalar field (EMCSF) and cosmological term will be a
solution of the Einstein frame scalar-tensor-MCSF equations with
cosmological term and with vanishing $\varphi$ or $\sigma$.
Therefore, we can construct Barker counterparts to any solution
of the EMCSF equations using the nonlinear action of the group
$SU(2)$.

An arbitrary element $U \in SU(2)$ can be presented in the form

\begin{equation}
U = {\cal D}(\delta){\cal O}(\beta){\cal D}(\gamma)
\end{equation}

where

\begin{equation}
{\cal D}(\gamma) = \pmatrix{e^{i\gamma/2} & 0 \cr 0 &
e^{-i\gamma/2} }
\end{equation}

and

\begin{equation}
{\cal O}(\beta) = \pmatrix{ \cos(\beta/2) & \sin(\beta/2) \cr -
\sin(\beta/2) & \cos(\beta/2)} \in SO(2) .
\end{equation}

It should be noted that the transformations (\ref{SUT})
associated with the subgroup consisting of the matrices ${\cal D}$
correspond to a constant shift of the scalar field $\sigma$:
$\sigma \rightarrow \sigma + const$. That is why, without loss of
generality, we may restrict ourselves to the subgroup $SO(2)$.

Taking as a seed solution $z_{0} = e^{i\sigma_{0}}$ and
performing a $SO(2)$ transformation with the  matrix  ${\cal
O}(\beta)$ we obtain a new solution $z$ which in terms of
$\varphi$ and $\sigma$ reads

\begin{equation}
\varphi = \arcsin\left[ \sin(\beta)\cos(\sigma_{0})\right] ,
\end{equation}

\begin{equation}\label{BSIG}
\sigma = \arcsin\left[ {\sin(\sigma_{0}) \over \sqrt{1 -
\sin^2(\beta)\cos^2(\sigma_{0})} } \right] \,\,\, .
\end{equation}

Once having the solution in the Einstein frame it is easy to
recover the corresponding Jordan frame one. The results we
summarize in the following proposition:

{\bf Proposition:} {\em Let $(g_{\mu\nu}, \sigma_{0})$ be a
solution to the EMCSF equations with a cosmological term
$\Lambda$. Then $({\tilde g}_{\mu\nu}, \Phi, \sigma)$ form  a
solution to the Barker-MCSF equations with a cosmological
potential $U(\Phi)= 2\Lambda \Phi^2$ where $\sigma$ is given by
(\ref{BSIG}) and }

\begin{equation}\label{JFBGS}
\Phi^{-1}(\sigma_{0}) = 1 - \sin^2(\beta)\cos^2(\sigma_{0}) ,
\end{equation}

\begin{equation}\label{JFBM}
{\tilde g}_{\mu\nu}= \Phi^{-1}(\sigma_{0}) g_{\mu\nu} .
\end{equation}

Without going into detail we formulate the proposition  in terms
of a perfect fluid:

{\bf Proposition:} {\em Let $(g_{\mu\nu}, \rho_{0},
u^{(0)}_{\mu})$ be a solution to the Einstein equations with an
irrotational stiff perfect fluid and a cosmological term
${\Lambda}$. Then $({\tilde g}_{\mu\nu}, \Phi, {\tilde \rho},
{\tilde u}_{\mu})$ is a solution to the Barker's equations with an
irrotational stiff perfect fluid and cosmological potential
$U(\Phi)= 2\Lambda \Phi^2$ if $\Phi$ and ${\tilde g}_{\mu\nu}$
are given by (\ref{JFBGS}) and (\ref{JFBM}), and }

\begin{eqnarray}\label{BTPFP}
{\tilde \rho}&=& \cos^2(\beta) \Phi^3(\sigma_{0}) \rho_{0} ,\nonumber \\
{\tilde u}_{\mu} &=& \Phi^{-1/2}(\sigma_{0}) u^{(0)}_{\mu} , \\
8\pi G_{*} \rho_{0} &=& - g^{\mu\nu}\partial_{\mu}\sigma_{0}
\partial_{\nu}\sigma_{0} .\nonumber
\end{eqnarray}

\subsection{Brans-Dicke theory}

Brans-Dicke theory is characterized by the functions $F(\Phi)=
\Phi$ and $Z(\Phi)= \omega/\Phi$ which correspond to ${\cal
A}^2(\varphi)= e^{2\alpha\varphi}$ where  $\alpha^2=\mid K\mid =
1/(3 + 2\omega)$.  In the Brans-Dicke case the metric
(\ref{DMSM}) takes the form

\begin{equation}\label{BDDMSM}
dl^2 = d\varphi^2 + e^{2\alpha \varphi}d\sigma^2 .
\end{equation}

The metric (\ref{BDDMSM}), up to the parameter $\alpha$, also
appears in the so called dilaton-axion gravity \cite{LWC}. It
should be noted, however, that we consider (\ref{BDDMSM}) in a
different physical context.  The symmetries of dilaton-axion
sector are well known, but for completeness we will describe them
again in our context.

Introducing the complex field

\begin{equation}
\xi= \alpha \sigma + i e^{-\alpha \varphi},
\end{equation}

we can write the metric (\ref{BDDMSM}) in the  Klein form

\begin{equation}
dl^2 = -{4\over \alpha^2} {d\xi d{\bar \xi}\over (\xi-{\bar
\xi})^2}.
\end{equation}

The dilaton-matter sector is $SL(2,R)$ invariant:

\begin{equation}\label{SLT}
\xi\rightarrow {\xi}^{{\,}\prime} = {A\xi + B\over C\xi + D}
\end{equation}

where  $$L = \pmatrix{A& B \cr C & D}\in SL(2,R).$$

There is also a discrete symmetry $\xi\rightarrow - {\bar \xi}$
which corresponds to $\sigma \rightarrow -\sigma$.

The Jordan frame metric ${\tilde g}_{\mu\nu}$ transforms under
(\ref{SLT}) as ${\tilde g}_{\mu\nu}^{{\,}\prime}
\Phi^{{\,}\prime}= {\tilde g}_{\mu\nu}\Phi $.

Let us consider a solution $(g_{\mu\nu}, \sigma_{0})$ to the EMCSF
equations. This solution is also a solution to the Einstein
frame  scalar-tensor equations with $\xi_{0}= i
e^{-\alpha\sigma_{0}}$. The $SL(2,R)$ transformation with the
matrix $L$ gives a new solution $z=\alpha\sigma + i
e^{-\alpha\varphi}$ where

\begin{equation}
e^{\alpha\varphi} = C^2 e^{-\alpha\sigma_{0}} +
D^2e^{\alpha\sigma_{0}} ,
\end{equation}

\begin{equation}\label{BDSIG}
\sigma = {1\over \alpha} {AC e^{-\alpha\sigma_{0}} + BD
e^{\alpha\sigma_{0}}
 \over  C^2 e^{-\alpha\sigma_{0}} + D^2 e^{\alpha\sigma_{0}} } .
\end{equation}

In this way we proved the following proposition:

{\bf Proposition:} {\em Let $\left(g_{\mu\nu}, \sigma_{0}\right)$
be a solution to the EMCSF equations with a cosmological term
$\Lambda$. Then $({\tilde g}_{\mu\nu}, \Phi, \sigma)$  form a
solution to the Brans-Dicke equations with a MCSF and a
cosmological potential $U(\Phi)= 2\Lambda \Phi^2$ where $\sigma$
is given by (\ref{BDSIG}) and }

\begin{equation}\label{JFBDGS}
\Phi^{-1/2}(\sigma_{0}) = C^2 e^{-\alpha\sigma_{0}} + D^2
e^{\alpha\sigma_{0}} ,
\end{equation}

\begin{equation}\label{JFBDM}
{\tilde g}_{\mu\nu} = \Phi^{-1}(\sigma_{0})g_{\mu\nu} .
\end{equation}

Applied to the case of a stiff perfect fluid the proposition
becomes:

{\bf Proposition:} {\em Let $(g_{\mu\nu},\rho_{0},
u^{(0)}_{\mu})$ be a solution to the Einstein equations with an
irrotational stiff perfect fluid and a cosmological term
$\Lambda$. Then $({\tilde g}_{\mu\nu}, \Phi,{\tilde \rho},
{\tilde u}_{\mu})$ is a solution to the Brans-Dicke equations
with  an irrotational stiff perfect fluid and a cosmological
potential $U(\Phi)=2\Lambda \Phi^2$   if $\Phi$ and ${\tilde
g}_{\mu\nu}$  are given by (\ref{JFBDGS}) and (\ref{JFBDM}), and }

\begin{eqnarray} \label{BDPFGF}
{\tilde \rho} &=& 4C^2D^2 \Phi^{3}(\sigma_{0}) \rho_{0}, \nonumber \\
{\tilde u}_{\mu} &=& \Phi^{-1/2}(\sigma_{0}) u^{0}_{\mu}, \\
8\pi G_{*} \rho_{0} &=& -
g^{\mu\nu}\partial_{\mu}\sigma_{0}\partial_{\nu}\sigma_{0} .
\nonumber
\end{eqnarray}

The transformations (\ref{SLT}) associated with the subgroup
consisting of matrices of the form

$${\cal T}= \pmatrix{A & B \cr 0 & A^{-1} }$$

correspond to either a constant shift or rescaling the
scalar-fields. That is why, without loss of generality, we may
restrict ourselves to the matrices

\begin{equation}
{\cal O}(\beta) =\pmatrix{ \cos(\beta/2)  & \sin(\beta/2) \cr
-\sin(\beta/2) & \cos(\beta/2)} \in SO(2).
\end{equation}

\subsection{${\cal A}(\varphi)\sim \cosh(\sqrt{\mid K\mid} \varphi)$ theory}

Here we consider the scalar-tensor theories with ${\cal
A}(\varphi)= \Omega \cosh(\sqrt{\mid K\mid}\varphi)$ , $\Omega >
0$. Introducing the complex field
\begin{equation}
\zeta = \exp(\Omega\sqrt{\mid
K\mid}{\,}\sigma)\left(\tanh(\sqrt{\mid K\mid }\varphi) + {i\over
\cosh(\mid K\mid \varphi)} \right)
\end{equation}

the metric can be written in the Klein form

\begin{equation}
dl^2 = -{4\over \mid K\mid} {d\zeta d{\bar \zeta}\over
(\zeta-{\bar \zeta})^2}.
\end{equation}

The case under consideration is analogous to the Bran-Dicke case
and we present directly the final results omitting the
intermediate details:

{\bf Proposition:} {\em Let $\left(g_{\mu\nu}, \sigma_{0}\right)$
be a solution to the EMCSF equations with a cosmological term
$\Lambda$. Then $({\tilde g}_{\mu\nu}, \Phi, \sigma)$  form a
solution to the equations of the ${\cal A}(\varphi)=\Omega
\cosh(\sqrt{\mid K\mid} \varphi)$ theory with a MCSF and a
cosmological potential $U(\Phi)= 2\Lambda \Phi^2$ if $\sigma$,
$\Phi$ and ${\tilde g}_{\mu\nu}$ are given by}

\begin{equation}
e^{2\Omega\sqrt{\mid K\mid}\, \sigma} = {1 + \left(
ACe^{\sqrt{\mid K\mid}\sigma_{0}} + BD e^{-\sqrt{\mid
K\mid}\sigma_{0}}  \right)^2  \over  \left( C^2 e^{\sqrt{\mid
K\mid}\sigma_{0}} + D^2 e^{-\sqrt{\mid K\mid}\sigma_{0}}
\right)^2 } ,
\end{equation}

\begin{equation}
\Phi^{-1}(\sigma_{0}) =\Omega^2 \left[1 + \left(ACe^{\sqrt{\mid
K\mid}\sigma_{0}} + BD e^{-\sqrt{\mid K\mid}\sigma_{0}} \right)^2
\right] ,
\end{equation}

\begin{equation}
{\tilde g}_{\mu\nu} = \Phi^{-1}(\sigma_{0})g_{\mu\nu} .
\end{equation}

It is not difficult to rewrite this proposition for  the case of
a stiff perfect fluid. The fluid energy density is given by

\begin{equation}
{\tilde \rho}= \Omega^2(AD + BC)^2\Phi^3(\sigma_{0})\rho_{0}.
\end{equation}

\subsection{Theory with a flat dilaton-matter sector}

As we have seen, the theory with a flat dilaton-matter sector is
described by ${\cal A}(\varphi)= 1  + a\varphi$. Let us introduce
the complex field

\begin{equation}
\varsigma = (1 + a\varphi)e^{i\sigma}.
\end{equation}

The dilaton-matter sector metric then takes the standard flat form

\begin{equation}
dl^2 = d\varsigma d{\bar \varsigma}.
\end{equation}

The metric is invariant under the group $Iso(\mathbf{R}^2)$ i.e.
under the transformations

\begin{equation}
\varsigma \rightarrow \varsigma^{\prime}= e^{i\vartheta}\varsigma
+ \eta
\end{equation}

where $\vartheta \in \mathbf{R}$ and $\eta = A + iB \in
\mathbf{C}$.

Let us take as a seed solution $\varsigma_{0} = 1 + a\sigma_{0}$.
It is not difficult to see that two of the parameters of the group
$Iso(\mathbf{R}^2)$ can be absorbed as constant shifts $\sigma
\rightarrow \sigma + const$, and only one essential parameter
remains. Performing a translation with $\eta = iB$ we obtain a new
solution $\varsigma$ with

\begin{equation}
(1+ a\varphi)= \sqrt{(1+ a\sigma_{0})^2 + B^2},
\end{equation}

\begin{equation}\label{SIGFDMS}
\sigma = {1\over a} \arcsin\left[{B\over \sqrt{(1+a\sigma_{0})^2 +
B^2}}\right]
\end{equation}

So we have:

{\bf Proposition:} {\em Let $\left(g_{\mu\nu}, \sigma_{0}\right)$
be a solution to the EMCSF equations with a cosmological term
$\Lambda$. Then $({\tilde g}_{\mu\nu}, \Phi, \sigma)$  form a
solution to the equations of the ${\cal A}(\varphi)=1 + a\varphi$
theory with a MCSF and a cosmological potential $U(\Phi)= 2\Lambda
\Phi^2$ if $\sigma$ is given by (\ref{SIGFDMS}) and  }

\begin{equation}
\Phi^{-1}(\sigma_{0}) = (1+ a\sigma_{0})^2 + B^2 ,
\end{equation}

\begin{equation}
{\tilde g}_{\mu\nu} = \Phi^{-1}(\sigma_{0})g_{\mu\nu}.
\end{equation}

In the case of a stiff perfect fluid we obtain

\begin{equation}
{\tilde \rho}= B^2 \Phi^{3}(\sigma_{0})\rho_{0}.
\end{equation}

\subsection{Other theories}

When the metric (\ref{DMSM}) has nontrivial isometries we can use
them to generate new solutions from known ones. In the general
case, however, the metric (\ref{DMSM}) has only trivial symmetry
$\sigma\longrightarrow \sigma + constant$. That is why the
solution generating procedure considered above fails.
Nevertheless, when the dilaton-matter sector does not possess
nontrivial symmetries we can still generate scalar-tensor
solutions starting with solutions to the EMCSF equations. For this
purpose we consider the geodesics of the metric (\ref{DMSM}). It
is not difficult to see that if $(\varphi(\tau), \sigma(\tau))$
is an affinely parameterized  geodesic and $(g_{\mu\nu},
\sigma_{0})$ is  a solution to the EMCSF equations with a
cosmological term $\Lambda$, then $(g_{\mu\nu},
\varphi(\tau)\mid_{\tau=\sigma_{0}},
\sigma(\tau)\mid_{\tau=\sigma_{0}})$ is a solution to the
scalar-tensor-MCSF equations with a cosmological term $\Lambda$.

In this way we obtain the  following

{\bf Proposition:} {\em Let $(\varphi(\tau), \sigma(\tau))$ is an
affinely parameterized geodesic of the metric (\ref{DMSM}) and
$(g_{\mu\nu}, \sigma_{0})$ is a solution  to the EMCSF equations
with a cosmological term $\Lambda$. Then $({\tilde g}_{\mu\nu},
\Phi, \sigma)$ form a solution to the scalar-tensor equations with
a MCSF and a cosmological potential $U(\Phi)=2\Lambda F^2(\Phi)$
where}

\begin{equation}
\sigma = \sigma(\tau)\mid_{\tau = \sigma_{0}},
\end{equation}

\begin{equation}
F^{-1}(\Phi(\sigma_{0})) = {\cal
A}^2(\varphi(\tau))\mid_{\tau=\sigma_{0}},
\end{equation}

\begin{equation}
{\tilde g}_{\mu\nu} = F^{-1}(\Phi(\sigma_{0}))g_{\mu\nu} .
\end{equation}

The geodesic equations for the metric (\ref{DMSM}) can be
formally solved and we obtain

\begin{eqnarray}
\int {d\varphi \over \sqrt{1 - C^2/{\cal A}^2(\varphi)}} = C_{1}
\pm
\tau  \,\,\, , \nonumber \\
\sigma  = C_{2} + C \int {d\tau \over {\cal A}^2(\varphi(\tau))}
\end{eqnarray}

where $C$, $C_{1}$ and $C_{2}$ are constants. The constants
$C_{1}$ and $C_{2}$ are unimportant and we shall omit them.

As a concrete example we consider the theory with $F(\Phi)= \Phi$
and $Z(\Phi)= {1\over 2}{\Phi^2 - 3\Phi + 3\over \Phi (\Phi -1)}$
corresponding to ${\cal A}^2(\varphi)= 1/(1+ \varphi^2)$. Solving
the geodesic equations for the metric $dl^2 = d\varphi^2 +
{d\sigma^2\over 1 + \varphi^2}$ we obtain

\begin{equation}
\varphi = {1\over C }\sqrt{1- C^2}\sin(C\sigma_{0}) \,\,\, ,
\end{equation}

\begin{equation}\label {OTS}
\sigma = {1 + C^2\over 2C}\sigma_{0} - {1-C^2\over
4C^2}\sin(2C\sigma_{0})
\end{equation}

where $0<C^2<1$.

Therefore, if $(g_{\mu\nu}, \sigma_{0})$ is a solution to the
EMCSF equations with a cosmological term $\Lambda$,  then
$({\tilde g}_{\mu\nu}, \Phi, \sigma)$ is a solution to the
equations of ${\cal A}^2(\varphi)= 1/(1 + \varphi^2)$-theory with
a MCSF and a cosmological potential $U(\Phi)=2\Lambda\Phi^2$ where
$\sigma$ is given by (\ref{OTS}) and

\begin{eqnarray}\label{JFOT}
\Phi(\sigma_{0}) &=& 1 + {1-C^2\over C^2}\sin^2(C\sigma_{0}), \nonumber \\
{\tilde g}_{\mu\nu} &=& \Phi^{-1}(\sigma_{0}) g_{\mu\nu} .
\end{eqnarray}

In the case of a stiff perfect fluid this solution corresponds to
the energy density

\begin{equation}\label{JFOTED}
{\tilde \rho} = C^2 \Phi^3(\sigma_{0}) \rho_{0}.
\end{equation}

\section{Examples of exact solutions}

As an illustration of the general techniques we consider some
particular examples of exact solutions. Here we set $\Lambda=0$.

\subsection{Homogeneous cosmological solutions}

Our first examples are homogeneous isotropic cosmological
solutions with an irrotational stiff perfect fluid. Consider the
metric

\begin{equation}
ds_{0}^2 = - dt^2 + t^{2/3}\left(dx^2 + dy^2 + dz^2\right) \,\,\,.
\end{equation}

This is a flat Friedman-Robertson-Walker metric for stiff perfect
fluid. The fluid energy density is $8\pi G_{*}\rho_{0} = {1/3t^2}$
and the 4-velocity is $u_{\mu} = -\partial_{\mu}t$. The fluid
velocity potential is $\sigma_{0}= - (1/\sqrt{3})\ln(t)$.

Since our seed solution is fixed the generated solutions will be
completely determined only by the explicit form of
$\Phi(\sigma_{0})$. That is why, for brevity, we shall present
only the explicit form of $\Phi(\sigma_{0})$ and the energy
density.

\subsubsection{Barker's theory}

Using the solution generating procedure for Barker's theory  we
obtain

\begin{eqnarray}\label{BHCS}
\Phi^{-1}(t) &=& 1 - \sin^2(\beta)\cos^2({1\over \sqrt{3}}\ln(t)),
\nonumber  \\
8\pi G_{*}{\tilde \rho} &=& {\cos^2(\beta)\over 3t^2
}\,\Phi^{3}(t) .
\end{eqnarray}

\subsubsection{Brans-Dicke theory}

In the case of Brans-Dicke theory the transformations
(\ref{BDPFGF}) yield the following solution:

\begin{eqnarray}\label{BDHCS}
\Phi^{-1/2}(t) &=& C^2 t^{\alpha/\sqrt{3}} +  D^2
t^{-\alpha/\sqrt{3}} ,\nonumber \\
{\tilde \rho}&=& {4C^2D^2\over 3t^2}\Phi^{3}(t).
\end{eqnarray}

\subsubsection{${\cal A}^2(\varphi)= 1/(1 + \varphi^2)$-theory}

Solution generating formulas (\ref{JFOT}) and (\ref{JFOTED}) give

\begin{eqnarray}\label{OTHCS}
\Phi(t) &=& 1 + {1-C^2\over C^2}\sin^2\left({C\over
\sqrt{3}}\ln(t)
\right)  , \nonumber \\
8\pi G_{*} {\tilde \rho} &=& {C^2\over 3t^2 } \Phi^3(t) .
\end{eqnarray}

The solution (\ref{BDHCS}) (presented in coordinates different
from those used here) was derived by Gurevich et al \cite{GFR}
and by Mimoso and Wands \cite{MW1} using a completely different
method. Although, the solutions (\ref{BHCS}) and (\ref{OTHCS})
can be also derived by the methods developed in Refs. \cite{BMIM},
\cite{MW1} they have not been presented so far in an explicit
form. It is worth noting that the methods used here to derive the
homogeneous cosmological solutions are much more elegant and
powerful than the methods developed in the previous works on the
subject.

\subsection{Inhomogeneous cosmological solutions}

Our next examples are much more interesting. We present new
inhomogeneous stiff perfect fluid cosmological solutions with
everywhere regular curvature invariants. The gravitational scalar
and the energy density are everywhere regular, too.

We consider the following solution to the Einstein equations with
an irrotational stiff perfect fluid \cite{PD}, \cite{Giovannini}:

\begin{eqnarray}\label{DSS}
ds^2_{0} &=& \cosh^{2(1-b)}(2mt)\cosh^{4b(2b-1)}(mr)(-dt^2 + dr^2)
\nonumber
\\ &+&  \cosh^{2(1-b)}(2mt)\sinh^{2}(mr)\cosh^{2(1-2b)}(mr)d\phi^2  \nonumber \\
&+& \cosh^{2b}(2mt)\cosh^{4b}(mr)dz^2 ,
\end{eqnarray}

\begin{eqnarray}
\sigma_{0}(t) &=& \sqrt{b^2 - 1}\arctan\left[\sinh(2mt)\right] .
\nonumber
\end{eqnarray}

The energy density and the four-velocity are given by

\begin{eqnarray}
8\pi G_{*}\rho_{0} = \nonumber \\ 4m^2 (b^2 - 1)\cosh^{-2(2 -
b)}(2mt)\cosh^{-4b(2b - 1)}(mr) ,
\end{eqnarray}

\begin{eqnarray}
u^{(0)}_{\mu} &=& \cosh^{1 - b}(2mt)\cosh^{2b(2b-1)}(mr) .
\end{eqnarray}

Here, $m$ and $b$ ($b^2>1$)  are free parameters.

The solution is everywhere  regular and satisfies the global
hyperbolicity condition. The curvature invariants are  regular
for whole spacetime. In addition, the spacetime describing by
this solution admits two commuting spacelike Killing vectors
${\partial\over \partial z}$ and ${\partial\over
\partial \phi}$ which are mutually and hypersurface orthogonal.

\subsubsection{Barker's theory}

Using as a seed solution (\ref{DSS}) and applying the solution
generating formulas (\ref{BTPFP}) we obtain the following
Barker's theory  three parametric  solution:

\begin{eqnarray}
\Phi^{-1}(t) = 1- \sin^2(\beta)\cos^2\left(\sqrt{b^2 -
1}\arctan\left(\sinh(2mt)\right)  \right) \nonumber  ,
\end{eqnarray}

\begin{eqnarray}\label{BTNSS}
{\tilde \rho}= \cos^2(\beta)\Phi^3(t)\rho_{0} .
\end{eqnarray}

When $\cos(\beta)=0$ the solution becomes a vacuum scalar-tensor
solution. That is why we shall focus our attention on the more
interesting case $cos(\beta)\ne 0$. In this case the
gravitational scalar and energy density are everywhere regular in
space-time. Moreover, we have calculated the curvature invariants

$$ {\tilde I}_{1}= {\tilde C}_{\mu\nu\alpha\beta}{\tilde
C}^{\mu\nu\alpha\beta}, \,\, {\tilde I}_{2}= {\tilde
R}_{\mu\nu}{\tilde R}^{\mu\nu},\,\, {\tilde I}_{3} = {\tilde
R}^2$$

with respect to the metric ${\tilde g}_{\mu\nu}=
\Phi^{-1}(\sigma_{0})g_{\mu\nu}$ and found that they are
everywhere regular. In this sense the solution (\ref{BTNSS}) is
nonsingular in the case $\cos(\beta)\ne 0$.

\subsubsection{Brans-Dicke theory}

In the case of Brans-Dicke theory, the solution generating
formulas (\ref{BDPFGF}) give

\begin{eqnarray}\label{BDTNSS}
\Phi^{-1/2}(t) =
\cos^2(\beta/2)\exp\left(\alpha\sqrt{b^2-1}\arctan\left(\sinh(2mt)\right)
\right)   \nonumber \\ +
\sin^2(\beta/2)\exp\left(-\alpha\sqrt{b^2-1}\arctan\left(\sinh(2mt)\right)
\right) \nonumber ,
\end{eqnarray}

\begin{eqnarray}
{\tilde \rho}= \sin^2(\beta)\Phi^{3}(t)\rho_{0}.
\end{eqnarray}

In the solution (\ref{BDTNSS}) we have restricted ourselves to the
subgroup $SO(2)\subset SL(2,R)$.

The gravitational scalar and the energy density are everywhere
regular in space-time. The curvature invariants ${\tilde I}_{1}$,
${\tilde I}_{2}$ and  ${\tilde I}_{3}$ are also everywhere
regular. In this sense the space-time described by the solution
(\ref{BDTNSS}) is nonsingular.

\subsubsection{${\cal A}^2(\varphi)= 1/(1 + \varphi^2)$-theory}

Solution generating formulas (\ref{JFOT}) and (\ref{JFOTED})
applied  to the seed solution (\ref{DSS}) yield  the following
solution:

\begin{eqnarray}
\Phi(t) = 1+ {1-C^2\over C^2}\sin^2\left[C \sqrt{b^2 -
1}\arctan\left(\sinh(2mt)\right)  \right] \nonumber ,
\end{eqnarray}

\begin{eqnarray}\label{AFTNSS}
{\tilde \rho}= C^2\Phi^{3}(t)\rho_{0}\nonumber .
\end{eqnarray}

The solution (\ref{AFTNSS}) is nonsingular. The gravitational
scalar and the energy density and the curvature invariants
${\tilde I}_{1}$, ${\tilde I}_{2}$, ${\tilde I}_{3}$ are regular
everywhere.

Clearly the solution generating techniques developed in the
present paper allow us to construct inhomogeneous cosmological
solutions with everywhere regular curvature invariants in many
other scalar-tensor theories different from those considered
above.

\subsection{Static spherically symmetric solutions}

Finally we present examples of new static, spherically symmetric
solutions.

As a seed solution we take the Janis-Newman-Winicour solution
\cite{JNW} to the EMCSF equations:

\begin{eqnarray}\label{EFSSS}
ds_{0}^2 &=& - f^{2\lambda}_{0}dt^2 + f^{-2\lambda}_{0}dr^2 +
f^{2(1-\lambda)}_{0} r^2 \left(d\theta^2 + \sin^2{\theta}d\phi^2
\right) \,\,\, ,\nonumber \\
\sigma_{0} &=& \sqrt{1 - \lambda^2}\ln(f_{0})
\end{eqnarray}

where $f^2_{0}= 1- {2M\over r}$ and $0\le\lambda\le 1$.

When we have  asymptotically flat seed backgrounds it is natural
to consider the solution generating transformations preserving the
asymptotic flatness of the field configuration. It is not
difficult to see that the subgroup  consisting of matrices of the
form
$$U = {\cal D}({\pi\over 2}){\cal O}(\beta){\cal D}^{-1}({\pi\over 2})$$
(i.e., isomorphic to $SO(2)\subset SU(2)$) preserves the
asymptotics in the case of Barker's theory. In the case of
Brans-Dicke theory, the subgroup preserving the asymptotics is
$SO(2)\subset SL(2,R)$ consisting of the matrices ${\cal
O}(\beta)$.

\subsubsection{Barker's theory}

The solution generating transformations which preserve the
asymptotics yield the following  solution:

\begin{eqnarray}
\Phi^{-1}(f_{0}) =  1 -
\sin^2(\beta)\sin^2\left(\sqrt{1-\lambda^2}\ln(f_{0})\right) ,
 \\
\sigma = \arcsin\left[\cos(\beta) \sin\left(\sqrt{1 -
\lambda^2}\ln(f_{0})\right) \over \sqrt{1 -
\sin^2(\beta)\sin^2\left(\sqrt{1-\lambda^2}\ln(f_{0})
\right)}\right] \nonumber \, .
\end{eqnarray}

\subsubsection{Brans-Dicke theory}

The scalar-tensor image of (\ref{EFSSS}) in the case of
Brans-Dicke theory is

\begin{eqnarray}
\Phi^{-1/2}(f_{0}) =
\cos^2(\beta/2)f_{0}^{{\,}\alpha\sqrt{1-\lambda^2}} +
\sin^2(\beta/2)f_{0}^{- \,\alpha\sqrt{1-\lambda^2}} , \\
\sigma= {1\over 2\alpha} \sin(\beta)
{f_{0}^{{\,}\alpha\sqrt{1-\lambda^2}} -f_{0}^{-
\,\alpha\sqrt{1-\lambda^2}} \over
\cos^2(\beta/2)f_{0}^{{\,}\alpha\sqrt{1-\lambda^2}} +
\sin^2(\beta/2)f_{0}^{- \,\alpha\sqrt{1-\lambda^2}}}. \nonumber
\end{eqnarray}

\subsubsection{${\cal A}^2(\varphi)= 1/(1 + \varphi^2)$-theory}

The solution generating transformations for the case under
consideration  yield the solution:

\begin{eqnarray}
\Phi &=& 1 + {1-C^2\over
C^2}\sin^2(C\sqrt{1-\lambda^2}\ln(f_{0})) \,\,\, ,\ \\
\sigma &=& {1+ C^2\over 2C}\sqrt{1-\lambda^2}\ln(f_{0}) \nonumber \\
 &-& {1-C^2\over 4C^2}\sin(2C\sqrt{1-\lambda^2}\ln(f_{0})) \,\,\, .
\nonumber
\end{eqnarray}

We recall that the Jordan frame metric in the above three examples
is given by ${\tilde g}_{\mu\nu}= \Phi^{-1}(f_{0})g_{\mu\nu}$
where the explicit form of $\Phi(f_{0})$ depends on the particular
scalar-tensor theory.

In this section we have constructed different kinds of exact
scalar-tensor solutions in the presence of a stiff perfect fluid
or a minimally coupled scalar field. Using the solution
generating  methods developed in the present work we can, of
course, generate many more and much more complicated scalar-tensor
solutions taking as a seed solutions the known in the literature
solutions of EMCSF equations (for example see
\cite{CCM}-\cite{L}). The physical properties of every generated
class scalar-tensor solutions, however, require a separate
investigation.

\section{Conclusion}

In this paper we have presented a general method for generating
exact solutions in some scalar-tensor theories of gravity with a
MCSF or irrotational stiff perfect fluid and with a potential
term of a special kind  for the gravitational scalar. The method
is based on the symmetries of the dilaton-matter sector in the
Einstein frame. In the case of Barker's theory the dilaton-matter
sector possess a group of symmetry $SU(2)$. In the case of
Brans-Dicke and the theory with "conformal coupling" the
dilaton-matter sector is $SL(2,R)$-symmetric.

We have described an explicit algorithm for generating exact
scalar-tensor solutions starting from solutions to the EMCSF
equations with a cosmological term  by employing  the non-linear
action of the symmetry  group of the dilaton-matter sector.

We have also presented a general method for generating
scalar-tensor solutions with a MCSF from solutions of EMCSF
equations using the geodesics of the Riemannian metric
(\ref{DMSM}) in the general case, when the  metric (\ref{DMSM})
does not possess nontrivial symmetries.

As an illustration of the solution generating techniques explicit
exact solutions have also been constructed. In particular, we
have constructed inhomogeneous cosmological solutions with
everywhere regular curvature invariants.

The solution generating method can be generalized for
scalar-tensor-Maxwell gravity. Taking into account that the
Maxwell action (in four dimensions) is conformally invariant we
can immediately write the scalar-tensor-MCSF-Maxwell  action in
the Einstein frame

\begin{eqnarray}
S ={1\over 16\pi G_{*}} \int d^{4}x \left(R
-2g^{\mu\nu}\partial_{\mu}\varphi \partial_{\nu}\varphi  \nonumber \right.\\
\left. - 2A^2(\varphi) g^{\mu\nu}\partial_{\mu}\sigma
\partial_{\nu}\sigma - F_{\mu\nu}F^{\mu\nu} - \Lambda\right)
\,\,\, . \nonumber
\end{eqnarray}

As is seen the dilaton-scalar-field sector possesses the same
symmetries (when they exist) as in the case of absence of Maxwell
field and they can be employed to generate new solutions with
electromagnetic field from known ones. In the general case, when
the metric associated with the dilaton-MCSF sector may not possess
nontrivial symmetries we can construct exact scalar-tensor
solutions in the presence of electromagnetic field using the
geodesics of the dilaton-scalar-field sector metric.

\section*{Acknowledgments}
I would like to thank V. Rizov for his interest in this work and
stimulating discussions.

This work was partially supported by Sofia University Grant No
459/2001.

\end{document}